\documentclass{article}

\usepackage[OT1]{fontenc}

\PassOptionsToPackage{table}{xcolor}

\usepackage{iclr2027_conference,times}

\usepackage{amsmath,amsfonts,bm}

\def\eqref#1{equation~\ref{#1}}

\def\1{\bm{1}}

\DeclareMathAlphabet{\mathsfit}{\encodingdefault}{\sfdefault}{m}{sl}
\SetMathAlphabet{\mathsfit}{bold}{\encodingdefault}{\sfdefault}{bx}{n}

\newcommand{\Sec}[1]{\S~{#1}}
\newcommand{\Fig}[1]{Fig.~{#1}}
\newcommand{\Tbl}[1]{Tbl.~{#1}}
\newcommand{\App}[1]{App.~{#1}}
\newcommand{\Paragraph}[1]{\noindent\textbf{#1}}

\usepackage[table]{xcolor}
\usepackage{booktabs}

\usepackage{tikz}
\definecolor{figurestepgreen}{HTML}{2C7773}
\newcommand{\FigureStep}[1]{%
  \tikz[baseline=(step.base)]{%
    \node[circle,fill=figurestepgreen,text=white,inner sep=0pt,
      outer sep=0pt,minimum size=8pt,font=\sffamily\bfseries\scriptsize]
      (step) {#1};%
  }%
}

\definecolor{gainbg}{RGB}{232,244,234}   
\definecolor{bestbg}{RGB}{205,232,211}   
\definecolor{lossbg}{RGB}{250,232,232}   

\AtBeginDocument{%
  \setlength{\intextsep}{2pt}%
  \setlength{\textfloatsep}{6pt plus 1pt minus 1pt}%
  \setlength{\floatsep}{4pt plus 1pt minus 1pt}%
  \setlength{\dbltextfloatsep}{6pt plus 1pt minus 1pt}%
  \setlength{\dblfloatsep}{4pt plus 1pt minus 1pt}%
}
\newcommand{\CompactTableLayout}{%
  \small
  \setlength{\abovecaptionskip}{2pt}%
  \setlength{\belowcaptionskip}{2pt}%
  \setlength{\tabcolsep}{4pt}%
  \renewcommand{\arraystretch}{0.9}%
}
\makeatletter
\def\compact@tabletype{table}
\AddToHook{cmd/@floatboxreset/after}{%
  \ifx\@captype\compact@tabletype\CompactTableLayout\fi
}
\makeatother
\AddToHook{env/wraptable/begin}{\CompactTableLayout}
\AddToHook{env/wrapfigure/begin}{%
  \small
  \setlength{\abovecaptionskip}{2pt}%
  \setlength{\belowcaptionskip}{0pt}%
}

\usepackage{graphicx}
\usepackage{placeins}
\usepackage{float}
\usepackage{wrapfig}
\floatstyle{ruled}
\newfloat{algorithm}{tbp}{loa}
\floatname{algorithm}{Algorithm}
\floatstyle{plain}
\usepackage{hyperref}
\usepackage{url}
\hypersetup{
  pdftitle={MegaFlux: Skew-Resilient MoE Megakernels via Pipelined Expert Replication},
  pdfauthor={Jianzhu Yao, Siva Kumar Sastry Hari, Vignesh Balaji, Sana Damani, Insu Jang, Pramod Viswanath, Christos Kozyrakis},
  pdfsubject={Preprint},
  hidelinks
}

\title{MegaFlux: Skew-Resilient MoE Megakernels via Pipelined Expert Replication}

\author{%
Jianzhu Yao\textsuperscript{1}, Siva Kumar Sastry Hari\textsuperscript{2}, Vignesh Balaji\textsuperscript{2},\\
\textbf{Sana Damani\textsuperscript{2}, Insu Jang\textsuperscript{2}, Pramod Viswanath\textsuperscript{1}, Christos Kozyrakis\textsuperscript{2}}\\
\textsuperscript{1}Princeton University \quad \textsuperscript{2}NVIDIA\\
\texttt{\{jy0246,pramodv\}@princeton.edu}\\
\texttt{\{shari,vbalaji,sdamani,insuj,ckozyrakis\}@nvidia.com}
}

\iclrpreprintcopy

\begin{document}

\maketitle

\begin{abstract}
Mixture-of-experts (MoE) megakernels fuse expert-parallel communication with expert computation.
However, under fixed expert placement, routing skew creates GPU stragglers: overloaded GPUs determine layer latency while others sit idle.
Replicating hot experts can shift work to underloaded GPUs, but dynamic replicas introduce additional work: replicas must receive expert weights to execute and, during training, their partial weight gradients must be reduced at the expert owners. 
We present \textbf{MegaFlux}, which makes expert replication a runtime decision and pipelines the communication induced by replication within persistent MoE execution.
An on-device planner jointly selects replica locations and assigns tile-aligned token blocks under a per-GPU replica budget, leaving router outputs unchanged.
The forward and backward megakernels realize pipelined expert replication: replicas begin computation as their required weights arrive, while backward overlaps replica-gradient reduction with ongoing expert computation.
MegaFlux extends TensorRT-LLM's CuTeDSL MegaMoE forward kernel and introduces a new backward MoE megakernel.
Across 147 configurations per direction on eight NVIDIA B200 GPUs,
MegaFlux achieves geometric-mean speedups of $1.45\times$ for forward
and $1.28\times$ for backward over the same megakernels with fixed placement, peaking at $2.14\times$ and $2.64\times$.
In ablations, pipelining hides $56$--$76\%$ of replica-weight transfer cost in forward and $91$--$100\%$ of combined weight-transfer and replica-gradient-reduction cost in backward, yielding up to $13.2\%$ and $26.7\%$ additional layer-latency reductions over the same replication plans with these operations executed separately. Integrated into vLLM for DeepSeek-V4-Pro prefill, MegaFlux delivers $1.13$--$1.26\times$ median end-to-end speedups over fixed placement.
\end{abstract}


\section{Introduction}
\label{sec:intro}


\Paragraph{Scaling MoE with expert parallelism.}
Mixture-of-experts (MoE) has become a common scaling strategy in recent large language models, including DeepSeek-V3, Qwen3, GLM-4.5, and Kimi K2~\citep{deepseekv3,yang2025qwen3,zeng2025glm,team2025kimi}. By activating only a few experts per token, MoE increases parameter capacity without proportionally increasing per-token computation~\citep{shazeer2017moe}. A learned router selects a small set of feed-forward experts for each token, whose outputs are combined according to the routing weights. To scale MoE capacity across multiple GPUs, expert parallelism (EP) partitions experts across GPUs, with each GPU storing only a subset of the expert parameters and routed activations sent to the GPUs hosting their selected experts~\citep{gshard}. This distributed execution introduces two system challenges: communication for moving routed activations and load imbalance when input-dependent routing concentrates work on experts hosted by only a few GPUs.

\Paragraph{Megakernels pipeline communication under fixed placement.}
Conventional EP executes token dispatch, expert computation, and token return as separate stages. Recent MoE megakernels instead fuse them into persistent kernels, enabling fine-grained pipelining of token movement and expert computation~\citep{flashmoe,cutedslmegamoe,zheng2026uniep,mok}. However, expert ownership remains fixed: a GPU can execute an expert only if it holds that expert's weights. Megakernels make each GPU's assigned work run efficiently, but do not change where that work can execute across GPUs.

\Paragraph{Routing skew motivates dynamic expert replication.}
Input-dependent routing can concentrate work on a few experts, leaving some GPUs underutilized while the GPUs hosting hot experts determine layer latency.
Training-time balancing mechanisms, including auxiliary load-balancing objectives and auxiliary-loss-free balancing, reduce such realized skew~\citep{deepseekv3,yang2025qwen3,muennighoff2025olmoe}.
In our captured Qwen3 and OLMoE workloads, routing skew slows fixed-placement MegaMoE by $7.3$--$41.0\%$.
Runtime systems therefore replicate hot experts and shift part of their routed work to underloaded GPUs~\citep{fastermoe,flexmoe,llep,moonep,ultraep}.
But dynamic replication introduces new communication: replicas must receive expert weights to execute and, during training, their partial weight gradients must be reduced at the expert owners. 
UltraEP, for example, distributes replica weights before forward token dispatch and aggregates replica gradients after expert backward~\citep{ultraep}.
The key challenge is to realize dynamic replication within the fine-grained communication--computation pipeline of MoE megakernels.

\Paragraph{Our solution.}
We present \textbf{MegaFlux}, which plans and executes dynamic expert replication within persistent MoE kernels.
Given the router's output, an on-device planner jointly chooses which experts to replicate, replica locations, and tile-aligned token-block assignment under a per-GPU replica budget.
The plan preserves token counts, expert selections, and routing weights, avoiding quality trade-offs from altering or dropping routed work.


Efficiently executing this plan requires integrating replica-induced communication into the persistent schedule.
We exploit two observations. First, a replica need not be fully materialized before computation begins: forward can begin its input projection once the corresponding weights and an input block are ready, while output-projection weights are still being transferred. Second, early gradient reduction requires early production of the corresponding partials on \emph{all} GPUs executing the replicated expert. Our backward therefore prioritizes gradient production for replicated experts across all participating GPUs, then reduces completed gradient tiles at their owners while other tiles are still being computed. We also use otherwise idle intervals in the persistent pipeline for replica-weight transfers and low-precision operand requantization, avoiding dedicated workers. Together, these mechanisms pipeline replica materialization and gradient reduction with tile-level expert computation, rather than executing them as separate stages around the megakernel.

\Paragraph{Evaluation.}
We build MegaFlux on TensorRT-LLM's CuTeDSL MegaMoE~\citep{cutedslmegamoe}, extending its forward kernel and introducing a new backward MoE megakernel.
On eight B200 GPUs spanning routing distributions and workload sizes, MegaFlux achieves geometric-mean forward/backward speedups of $1.45\times$/$1.28\times$ over fixed placement, peaking at $2.14\times$/$2.64\times$.
Ablations isolate the two effects: separate-stage replication reduces forward/backward latency by up to $42.4\%/51.2\%$, with pipelining providing a further $13.2\%/26.7\%$ reduction.
We also compare with Megatron-Core, UltraEP, and Mixture-of-Kittens in BF16. 
Integrated into vLLM for DeepSeek-V4-Pro, MegaFlux delivers $1.13$--$1.26\times$ end-to-end prefill speedups over fixed placement.

\Paragraph{Contributions.}

{\setlength{\parskip}{3pt} 

\noindent $\bullet$ We formulate block-granular runtime expert replication and develop an on-device planner for replica placement and token-block assignment under per-GPU replica capacity limits.

\noindent $\bullet$ We extend CuTeDSL MegaMoE forward with pipelined expert replication, using phase-level weight readiness to overlap replica materialization with expert computation.

\noindent $\bullet$  We design a backward MoE megakernel that pipelines data- and weight-gradient computation, requantization, token movement, replica-weight transfer, and gradient reduction within one persistent schedule.

}
\section{MegaMoE Execution and Motivation}
\label{sec:background-motivation}

\subsection{Fixed-Placement MegaMoE Execution}

\Paragraph{Expert-parallel MoE.}
Expert parallelism (EP) distributes $E$ experts across $P$ GPUs, one rank per GPU~\citep{gshard}.
Each expert's \emph{owner} stores its original weights.
Each rank initially holds $M$ input tokens, each routed to $k$ distinct experts; expert outputs return to their source ranks for weighted combination.
Let $\mathcal S_t$ and $p_{t,e}$ be token $t$'s selected experts and routing weights. Grouping the routed input rows as $X_e$, expert $e$ computes
\begin{equation}
  Z_e=X_eW_{1,e},\quad A_e=f(Z_e),\quad O_e=A_eW_{2,e},\qquad
  y_t=\sum_{e\in\mathcal S_t}p_{t,e}O_e[t],
  \label{eq:moe-forward}
\end{equation}
Here $Z_e=[G_e,U_e]$, $f(Z_e)=\operatorname{SiLU}(G_e)\odot U_e$, and
$O_e[t]$ denotes expert $e$'s output for token $t$. We refer to the two projections as FC1 and FC2. The equations omit quantization.

\Paragraph{Persistent MoE execution.}
Conventional EP implementations often execute token dispatch, expert computation, and token return as separate GPU kernels, introducing repeated launch and synchronization boundaries and limiting overlap across stages.
Persistent MoE megakernels instead keep cooperative thread arrays (CTAs) resident and use warp-specialized workers for scheduling, token movement, matrix multiply--accumulate (MMA), and epilogues.
Readiness flags and barriers expose tile-level producer--consumer dependencies, allowing communication and computation to proceed concurrently.
Our implementation targets Blackwell and uses \texttt{tcgen05} MMA with FP32 accumulation in tensor memory (TMEM)~\citep{cutlassblackwell}. Native block-scaled MMA supports MXFP4/MXFP8 computation, including FP4 weights with FP8 activations; the low-precision forward path further uses two-CTA MMA and a runtime MMA token dimension to reduce padding overhead. Tensor Memory Accelerator (TMA) stages operands into shared memory~\citep{cudahoppertuning}.

\Paragraph{Fixed-placement MegaMoE forward.}
Our forward baseline is TensorRT-LLM's CuTe DSL MegaMoE implementation~\citep{cutedslmegamoe}, which overlaps token transfers with tiled FC1/FC2 computation instead of using separate kernel stages.
As FC1's epilogue produces each complete activation block, the corresponding FC2 work can begin while other blocks remain in flight; output return begins after all FC2 blocks of the expert complete.
Metadata routing and the final top-$k$ sum remain separate launches.
This schedule overlaps communication with computation, but every expert tile still executes on its owner.


\Paragraph{Fixed-placement MegaMoE backward.}
We implement a fixed-placement backward megakernel for data gradients (DGrad) and weight gradients (WGrad).
Ignoring quantization, define $D_e[t]=p_{t,e}dy_t$ and let $dZ_e$ be the SwiGLU backward result from $(Z_e,dA_e)$. Then
\begin{equation}
  dA_e=D_eW_{2,e}^{\top},\quad dX_e=dZ_eW_{1,e}^{\top},\quad
  dW_{2,e}=A_e^{\top}D_e,\quad dW_{1,e}=X_e^{\top}dZ_e.
  \label{eq:moe-backward}
\end{equation}
The megakernel streams token blocks through DGrad, fusing SwiGLU backward and feature-axis quantization, while preparing token-axis-quantized $D_e$ and $dZ_e$ for $dW_{2,e}$ and $dW_{1,e}$ (DGrad and WGrad reduce along different dimensions). CTAs independently transition from DGrad to WGrad, allowing token return and requantization to overlap with ongoing expert computation; the final top-$k$ sum remains separate. 




\subsection{Routing Skew in Real Workloads}
\label{sec:routing-skew}
\makeatletter\@afterindenttrue\makeatother
\begin{wrapfigure}{R}{0.49\textwidth}
  \centering
  \includegraphics[width=\linewidth]{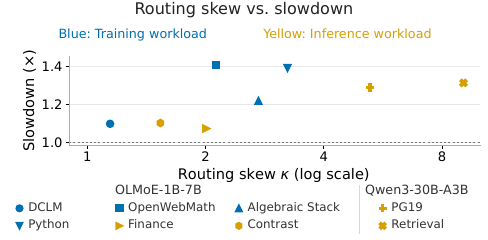}
  \caption{Routing skew increases fixed-placement MegaMoE latency. Points correspond to the P95-skew routes from OLMoE-1B-7B and Qwen3-30B-A3B workloads. Slowdowns are relative to an equal-token balanced route.}
  \label{fig:motivation}
  \label{fig:routing-skew}
  \label{fig:trace-skew}
  \label{fig:skew-sensitivity}
\end{wrapfigure}

\Paragraph{Routing profiles.}
We collect expert-routing profiles from Qwen3-30B-A3B~\citep{yang2025qwen3}
serving PG19 and InfiniteBench retrieval~\citep{rae2019compressive,zhang2024bench},
as well as from OLMoE-1B-7B~\citep{muennighoff2025olmoe} during training on DCLM,
OpenWebMath, Algebraic Stack, and StarCoder Python
~\citep{li2024datacomp,paster2024openwebmath,azerbayev2024llemma,li2023starcoder}.
We also use OLMoE inference routes on Finance and Contrast~\citep{trace_finance}.
These workloads span inference and training with varied expert-load imbalance.

\Paragraph{Quantifying routing skew.}
For each layer invocation or recorded-token window, let $c_e$ denote the number of tokens routed to expert $e$ among $E$ experts.
Following DA-MoE~\citep{huang2026decoding}, we measure routing skew as
\[
\kappa=E\sum_{e=1}^{E}\left(\frac{c_e}{\sum_j c_j}\right)^2.
\]
Uniform routing gives $\kappa=1$; larger values indicate more uneven expert loads.

\Paragraph{Latency impact under fixed placement.}
\Fig{\ref{fig:routing-skew}} evaluates the 95$^{th}$-percentile-skew (P95-skew) routing profile from each workload using CuTeDSL MegaMoE with the source model's native MoE configuration and trained checkpoint weights.
We use BF16 on eight B200 GPUs (EP8), with 8K tokens/rank for Finance and Contrast and 16K for the other six datasets.
We replay captured hidden states and routing weights and compare against a balanced route that changes only the expert IDs while preserving the token count.
Across the OLMoE and Qwen3 workloads, these P95 routes incur $7.3$--$41.0\%$ higher latency than their balanced counterparts. Thus, even when communication and expert computation are already pipelined within MegaMoE, fixed expert placement leaves substantial latency exposed to input-dependent routing skew.
\section{MegaFlux}
\label{sec:method}

An on-device planner converts the current routing into a block-granular replica placement and work assignment (\Sec{\ref{sec:planner}}). The executor realizes this plan inside the forward and backward megakernels, pipelining replica-weight transfers and gradient reduction with expert computation (\Sec{\ref{sec:method-rebalancing-fusion}}).

\subsection{Planner: Replica Placement and Token-Block Assignment}
\label{sec:planner}
For each MoE layer, given the current routing profile, the planner places expert \emph{replicas} across available GPUs and assigns tile-aligned token blocks subject to a per-GPU replica budget. This optimization leaves expert selections and routing weights unchanged.

\Paragraph{Problem formulation.}
Splitting token counts among copies can create a partial tile at each copy.
We instead partition each expert's routed rows into \emph{token blocks} of at most $\beta$ rows, matching the token dimension of the megakernel's tiled FC1/FC2 computation, leaving at most one partial block per expert regardless of replication. The same assignment determines which GPU processes each DGrad row and which copies contribute partial WGrad tiles.

Let $c_e$ be the number of tokens routed to expert $e$ across all GPUs, $h(e)$ its owner, and $b_e=\lceil c_e/\beta\rceil$ its block count. Let $q_{er}$ denote the number of expert $e$'s blocks assigned to GPU $r$, and $s$ the replica budget per GPU. Since experts share dimensions, tile-aligned block counts provide a lightweight proxy for compute load. We model the compute load of GPU $r$, and seek to balance this load subject to the replica-capacity constraint:
\begingroup
\setlength{\abovedisplayskip}{4pt}
\setlength{\belowdisplayskip}{4pt}
\begin{equation}
\begin{alignedat}{2}
\min_{q\in\mathbb{Z}_{\geq 0}^{E\times P}}\quad
& \max_r \ell_r,
&\qquad& \ell_r=\sum_e q_{er},\\
\text{s.t.}\quad
& \sum_r q_{er}=b_e\quad(\forall e),
&& \sum_{e:h(e)\ne r}\mathbf{1}[q_{er}>0]\le s\quad(\forall r).
\end{alignedat}
\label{eq:planner-blocks}
\end{equation}
\endgroup
For $r\ne h(e)$, $q_{er}>0$ requires a replica. $q$ jointly specifies replica placement and block assignment. 

\Paragraph{Bounded greedy search.}
MegaFlux runs four greedy searches in parallel. Each starts from the no-replica assignment and shifts blocks from overloaded owners toward $\tau=\lceil\sum_e b_e/P\rceil$, using existing replicas before opening new copies on the least-loaded eligible GPUs.

The searches combine donor priority (decreasing total overload or largest remaining home-expert load) with fan-out (at most one new copy per expert per round, or multiple copies). Each round visits overloaded owners once. A donor visit ends when its load falls to $\tau$ or below or its expert scan finishes; an expert's fan-out stops when its local blocks are exhausted or no eligible receiver remains. Each search stops after $s+2$ rounds or a round with no block movement. We minimize maximum GPU load, breaking ties by replica count, then prune redundant replicas and refine assignments without increasing that maximum. This defines \emph{load-only planning} (\App{\ref{app:planner-search}}).
For backward, we use a \emph{reduction-aware} variant that additionally accounts for the owner-side cost of replica-gradient reduction. During pruning, it uses the fixed proxy $\max_r(\ell_r+\lambda n_r)$, where $n_r$ counts remote replicas of experts owned by GPU $r$. We fix $\lambda=12$ for all reported backward configurations (\App{\ref{app:planner-cost}}).


\Paragraph{On-device planning and route construction.}
Four warps evaluate the four candidates independently in parallel.
After selecting the best candidate, the planner warps cooperatively materialize a block-to-GPU table from $q$. A stable ordering within each expert gives every routed assignment a block index that directly selects its destination GPU.

\par
\ifnum\value{WF@wrappedlines}>1
  \vspace{\dimexpr\baselineskip*\numexpr\value{WF@wrappedlines}-1\relax\relax}
\fi
\WFclear

\subsection{Executor: Pipelined Expert Replication}
\label{sec:method-rebalancing-fusion}
\begin{figure}
    \centering
    \includegraphics[width=\linewidth]{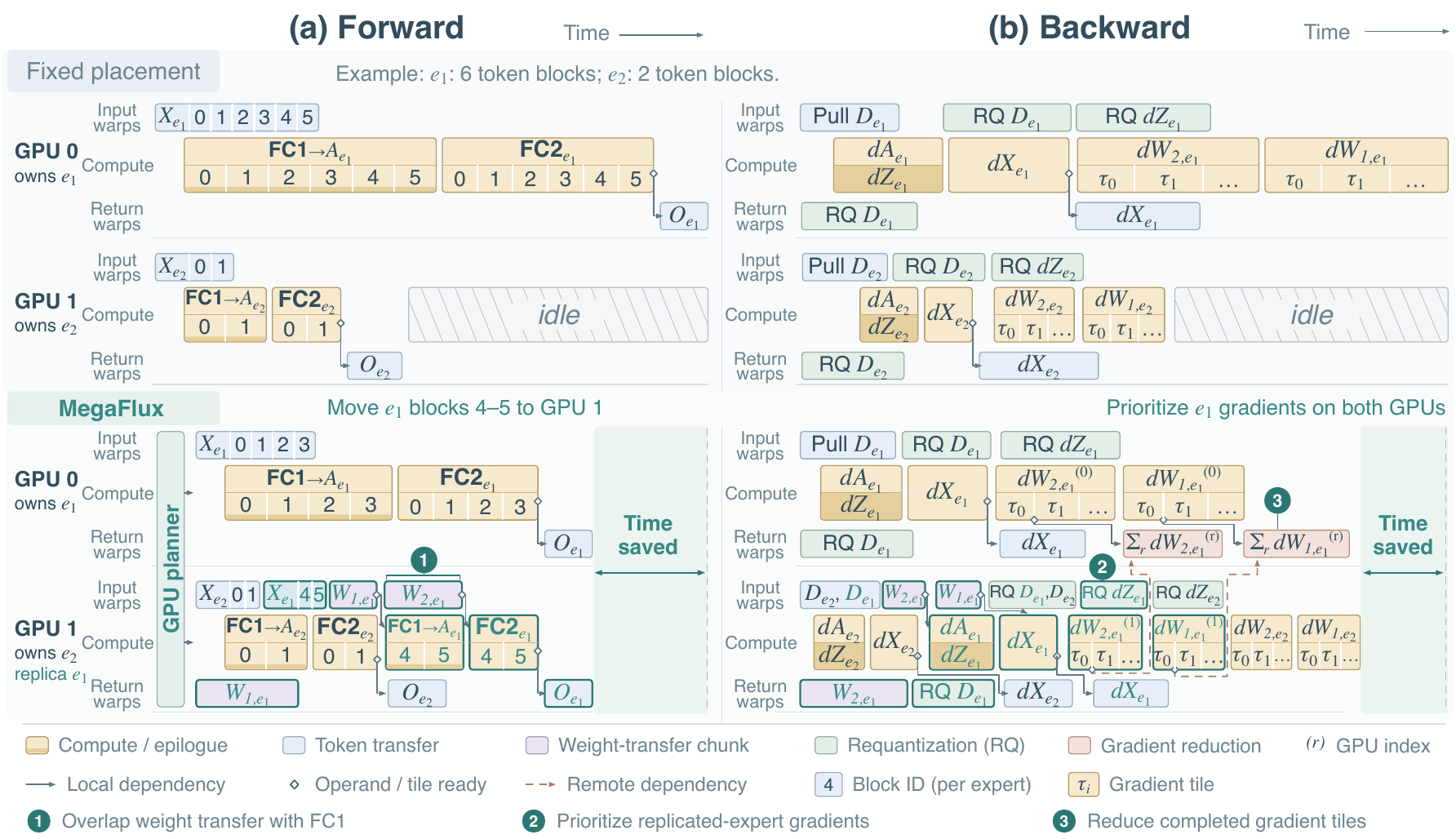}
    \vspace{-2em}
\caption{
\textbf{Pipelined expert replication in MegaFlux.}
We compare fixed placement MegaMoE (top) with MegaFlux (bottom), whose on-device planner redistributes token blocks across GPUs.
\emph{Forward (left):} separate $W_{1,e}$/$W_{2,e}$ readiness allows replica FC1 to overlap $W_{2,e}$ transfer and expert computation.
\emph{Backward (right):} replica-weight transfer and requantization ($\mathrm{RQ}$) overlap DGrad; prioritizing $e_1$'s weight-gradient
production on both GPUs enables earlier gradient reduction at $e_1$'s owner that overlaps remaining computation.
Timelines are schematic.
}
    \label{fig:megamoe}
\end{figure}

MegaFlux pipelines replica-weight transfer and, in backward, replica-gradient reduction with expert computation. Phase- and tile-level readiness lets each operation begin as soon as its dependencies are satisfied (\Fig{\ref{fig:megamoe}}).

\Paragraph{Forward.}
A replica can begin useful work before all of its weights arrive: FC1 requires only $W_1$ and an input token block, while FC2 requires $W_2$ and the completed activation block from FC1. Separate readiness for the two weight matrices therefore overlaps FC1 with $W_2$ transfer (\Fig{\ref{fig:megamoe}}, \FigureStep{1}), while owned-expert computation overlaps either transfer.
Operand loaders enforce these conditions within the existing tile schedule, without a replica-wide barrier.
Replica-weight movement reuses the communication warps rather than reserving a separate worker group. A small subset services the replica-weight queue from megakernel entry, while the remaining input-transfer warps first drain their assigned token blocks and then join the same queue.
This progressively increases the resources available for replica-weight transfer while preserving the existing token-movement pipeline.

\Paragraph{Backward.}
Backward overlaps both replica materialization and replica-specific gradient reduction with the existing DGrad--WGrad pipeline. DGrad consumes $W_{2,e}$ before $W_{1,e}$, so we apply the same per-projection readiness in the reverse order, overlapping both transfers with ongoing DGrad. Our MXFP8 backward additionally prepares token-axis $D_e$ and $dZ_e$ by requantization for WGrad because DGrad and WGrad reduce along different dimensions. We schedule this conversion in otherwise idle communication intervals, and each worker transitions independently from DGrad to WGrad after retiring its previous accumulator state, avoiding a device-wide phase barrier.

Overlapping replica-gradient reduction requires more than producing one partial early: the owner can reduce a gradient tile only after matching partials from all participating GPUs are ready. MegaFlux therefore prioritizes WGrad preparation and gradient-tile production for replicated experts across all participants (\Fig{\ref{fig:megamoe}}, \FigureStep{2}).
As soon as the required partials for a tile become available, owner-side workers pull and accumulate them in FP32 in a fixed order while other gradient tiles continue to be produced (\Fig{\ref{fig:megamoe}}, \FigureStep{3}). This moves the complete reduction dependency earlier and pipelines gradient reduction with the remaining expert computation.



\section{Evaluation}
\label{sec:evaluation}

We evaluate MegaFlux along three axes: the latency benefit of dynamic expert replication across routing skew and workload sizes, the benefit of pipelining replica operations inside the megakernel over executing the same replication plan in separate stages, and the cost and quality of online planning. We also compare MegaFlux against other existing MoE baselines. 

\subsection{Experimental Setup}
\label{sec:eval-setup}

\Paragraph{Platform and configurations.}
We evaluate on eight NVIDIA B200 GPUs within one NVLink domain (EP8). Each layer uses hidden/intermediate dimensions 7168/2048, SwiGLU, and top-$8$ routing. Each GPU owns $E/P$ experts and has capacity for $s=E/(4P)$ additional replicas, giving 25\% spare expert capacity; the planner uses $\beta=128$ token rows per block.
The main grid covers $E\in\{64,128,256\}$ and $M\in\{1,2,4,8,16,32,64\}$K input tokens per rank, where 1K is 1024. Forward uses MXFP4 weights and MXFP8 activations; backward uses MXFP8 operands, with FP32 accumulation and BF16 outputs in both. Our primary baseline is the same MegaMoE kernel with fixed expert placement. All variants pass numerical correctness checks. To enable a finer-grained controlled sweep across expert counts, workload sizes, and routing distributions, we use synthetic weights and activations while preserving trace-derived and synthetic Zipf expert-load distributions.

\begin{figure}
  \centering
  \includegraphics[width=\linewidth]{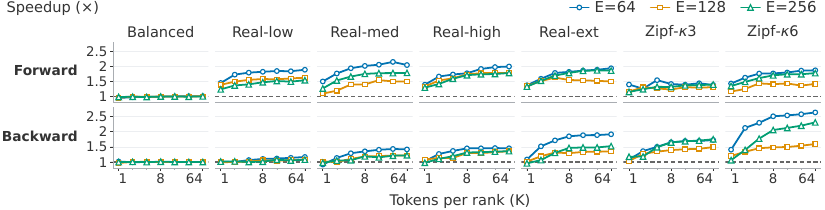}
  \vspace{-2.5em}
  \caption{MegaFlux speedup over fixed placement in the same megakernel on EP8. Forward uses MXFP4/MXFP8 and includes online planning; backward uses MXFP8. Columns show routing distributions and curves vary expert count. Trace-derived skew labels are source-relative.}
  \label{fig:eval-main}
\end{figure}

\Paragraph{Routing workload construction.}
\label{sec:method-benchmark}
We construct trace-derived routing workloads from Qwen3 retrieval histograms for forward and OLMoE training histograms for backward (\Sec{\ref{sec:routing-skew}}). From captures with 32K input tokens per rank, we select three histograms at each of four skew levels: \emph{real-low}, \emph{real-median}, \emph{real-high}, and \emph{real-extreme}.
We rebin their normalized expert loads to the target expert count and rescale to the target token count, enforcing $k$ distinct experts per token. We then permute expert loads across fixed owners and construct routes that realize the resulting counts. These workloads preserve observed expert-load distributions while allowing controlled sweeps over expert count and token count; they are not end-to-end replays of the source models. We add balanced routing ($\kappa=1$) and Zipf distributions $\rho_e(\alpha)\propto e^{-\alpha}$ ($\alpha$ calibrated to achieve $\kappa\approx3$ or $6$), a common controlled model for nonuniform expert popularity~\citep{tempo}. All methods evaluated use identical routes, routing weights, and initial expert ownership. \App{\ref{app:eval-workloads}} provides more details. 

\Paragraph{Measurement.}
We time forward and backward separately.
For the main comparisons, each measurement includes token transport, expert computation, output combination, and required replica-weight transfers; backward additionally includes replica-gradient reduction. We exclude the router and shared expert and report rank-maximum latency.
Forward uses load-only planning, with planner latency included. Backward reuses a reduction-aware plan selected before the corresponding training forward step; planning is charged once to that forward pass. On real-high training routes ($E=256$, 1K--32K tokens/rank), reduction-aware planning changes forward execution latency by just $-0.8\%$ to $+0.5\%$ versus load-only planning.

\subsection{Performance Relative to Fixed Placement}
\label{sec:eval-main}

\Paragraph{Dynamic expert replication reduces straggler latency.}
\Fig{\ref{fig:eval-main}} reports geometric-mean speedups of $1.45\times$ for forward and $1.28\times$ for backward over fixed placement.
Peak speedups (forward/backward) are $2.14\times$/$1.43\times$ for real-median, $1.99\times$/$1.46\times$ for real-high, and $1.93\times$/$1.92\times$ for real-extreme. All 84 Zipf settings 
improve, with peak forward/backward speedups of $1.53\times$/$1.75\times$ for Zipf-$\kappa$3 and $1.86\times$/$2.64\times$ for Zipf-$\kappa$6.
MegaFlux reduces the busiest GPU's excess token--expert assignments above the per-GPU average by a median 99.6\%/98.2\% in forward/backward (\App{\ref{app:assignment-imbalance}}). 

\Paragraph{Workload size determines the payoff.}
Gains generally strengthen as more computation amortizes planning and replica-operation overhead. For example, at $E=256$ under real-high routing, forward speedup
increases from $1.29\times$ at 1K to $1.77\times$ at 64K tokens per rank.
Balanced routes remain near parity ($0.985\times$ forward, $0.999\times$ backward geometric mean). Small skewed backward workloads can still regress to $0.91\times$. Thus, near-balanced assignments do not guarantee lower latency; exposed replica-operation costs may outweigh the benefit of redistributing work.

\begin{figure}
  \centering
  \includegraphics[width=\linewidth]{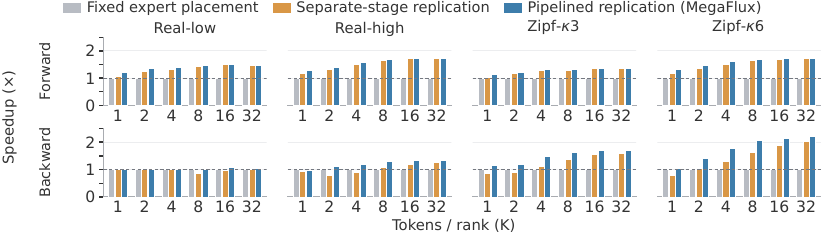}
  \vspace{-2.5em}
  \caption{Speedup over fixed placement for separate-stage and pipelined expert replication (EP8, $E=256$, top-$8$, 25\% spare expert capacity). Both use identical replica-placement and token-block-assignment plans. The separate-stage variant transfers replica weights before expert computation and, in backward, reduces replica gradients afterward; MegaFlux pipelines these replica operations with expert computation inside the megakernel.}
  \label{fig:eval-fusion-three-way}
\end{figure}

\subsection{Decomposing Pipelined Expert Replication}
\label{sec:eval-hidden-ratio}

\Paragraph{Rebalancing and pipelining provide complementary gains.}
\Fig{\ref{fig:eval-fusion-three-way}} compares separate-stage replication with MegaFlux using identical replica-placement and token-block-assignment plans. 
Across real-low/high and Zipf-$\kappa3/\kappa6$ at 1K--32K tokens/rank, separate-stage replication reduces latency by up to $42.4\%$ forward and $51.2\%$ backward; while pipelining provides an additional $13.2\%$ and $26.7\%$ reduction relative to the separate-stage variant.
For small workloads, pipelining is particularly effective: exposed replica-operation costs can outweigh the compute saved by rebalancing, but pipelining recovers these gains by hiding much of that overhead.
For backward Zipf-$\kappa6$ at 1K, pipelining turns a $27.4\%$ slowdown into a $5.6\%$ latency reduction versus fixed placement.


\begin{wrapfigure}{R}{0.5\textwidth}
  \centering
  \includegraphics[width=\linewidth]{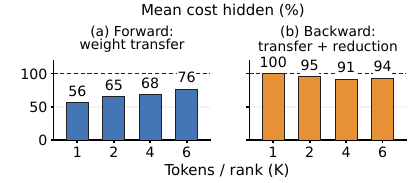}
  \vspace{-1.5em}
  \caption{Estimated replica-operation cost hidden by pipelining on real-high routes (EP8, $E=256$, top-$8$, 25\% spare capacity). Bars show the mean of two runs.}
  \label{fig:eval-hidden}
\end{wrapfigure}

\Paragraph{Workload size changes the relative contributions.}
More tokens amortize replica-operation costs over more computation.
In separate-stage replication, forward speedup grows from $1.03$--$1.17\times$ at 1K to $1.37$--$1.74\times$ at 32K, while the additional saving from pipelining falls from $9.6$--$13.2\%$ to $-0.3$--$0.2\%$.
Backward retains $3.4$--$7.0\%$ additional savings from pipelining at 32K, where replica-gradient reductions also overlap computation; for Zipf-$\kappa6$, pipelining raises speedup from $2.05\times$ to $2.20\times$.

\Paragraph{Hidden replica-operation cost.}
\Fig{\ref{fig:eval-hidden}} fixes real-high routes at 1K/2K/4K/6K tokens per rank.
The estimated hidden fraction is 56--76\% for forward weight transfer and approximately 91--100\% for backward weight transfer plus gradient reduction.
We estimate the hidden fraction as one minus the residual overhead under pipelined execution divided by the cost of executing the replica operations separately. 
The residual overhead is MegaFlux's excess latency over a same-plan baseline MegaMoE latency that excludes replica operations. The separate cost covers replica-weight transfer in forward, and replica-weight transfer plus replica-gradient reduction in backward. 
\WFclear

\subsection{MegaFlux compared to prior art}
\label{sec:eval-other-moe-baselines}

\begin{figure}
  \centering
  \includegraphics[width=\linewidth]{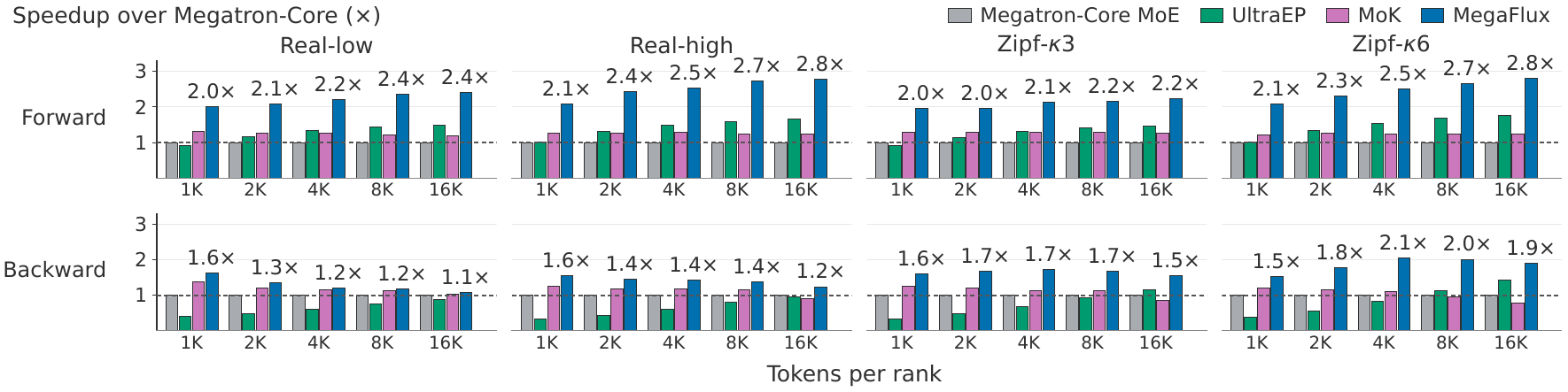}
  \vspace{-2.5em}
  \caption{BF16 layer speedup relative to Megatron-Core on EP8, $E=256$, with 25\% capacity for replicas. Rows show forward/backward; columns show routing distributions. Unmodified MoK additionally computes a shared expert and, in backward, shared-expert gradients and routing-coefficient gradients; the text reports comparable-work estimates.}
  \label{fig:eval-other-moe}
\end{figure}

We compare MegaFlux to Megatron-Core and UltraEP, both of which use DeepEP 2.1.0 and Transformer Engine 2.9.0 grouped GEMMs. We also compare to unmodified Mixture-of-Kittens 0.1.0 (MoK)~\citep{mok}. For a fair comparison, we implement a BF16 variant of MegaFlux and evaluate three other baselines in BF16; we evaluate $E=256$, 1K--16K tokens/rank, and four routing distributions (\Fig{\ref{fig:eval-other-moe}}). Timings include layer computation, planning, and replica operations; MegaFlux and UltraEP use the same replica budget. UltraEP's gradient reduction completes within the measured layer, without overlapping non-MoE work.


MegaFlux achieves geometric-mean forward/backward speedups of $2.31\times$/$1.53\times$ over Megatron-Core, $1.74\times$/$2.35\times$ over UltraEP, and $1.84\times$/$1.37\times$ over unmodified MoK. MoK additionally computes a shared expert and, in backward, its gradients and routing-coefficient gradients. For a comparable-work estimate, we benchmark this extra work with cuBLAS GEMMs and a fused routing-gradient kernel and add its cost serially to MegaFlux without assuming overlap. The adjusted speedups over MoK are $1.70\times$ forward and $1.26\times$ backward.

\subsection{Online Planner}
\label{sec:eval-planner}
\makeatletter\@afterindenttrue\makeatother
\begin{wrapfigure}{R}{0.5\textwidth}
  \centering
  \includegraphics[width=\linewidth]{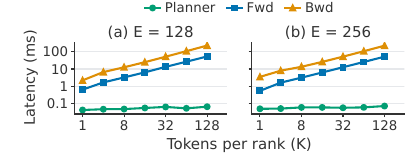}
  \caption{Planner and forward/backward latency on EP8, real-high routes (log scale).}
  \label{fig:eval-planner}
\end{wrapfigure}

\Paragraph{Planner cost.}
We evaluate planner cost with a real-high experiment with $E=128/256$ from 1K to 128K tokens per rank (\Fig{\ref{fig:eval-planner}}). 
Planning takes 42.6--65.9$\,\mu$s for $E=128$ and 49.8--72.0$\,\mu$s for $E=256$, growing much less than layer execution.
At $E=256$, the ratio of planning time to median forward latency decreases from 9.17\% to 0.14\% as tokens per rank increase from 1K to 128K.
Considering the sum of separately measured forward and backward median latencies, this ratio decreases from 1.24\% to 0.03\%.
Planning is therefore most visible at small token counts, whereas larger workloads amortize it over more expert computation.

\Paragraph{Plan quality.}
We compare our planner with UltraEP's ($E=256$, real-high routes, eight replica slots/rank).
First, we apply both planners' assignments to MegaFlux.
Across 1K--32K tokens/rank, our plans deliver $0.981$--$1.102\times$ the forward-plus-backward (F+B) performance of UltraEP's plans, remaining within 2\% beyond 1K.
Second, we apply both planners' assignments to the UltraEP baseline built on DeepEP communication and Transformer Engine grouped GEMMs.
At 1/4/8/16K tokens/rank, our plans deliver $2.21/1.37/1.07/0.99\times$ the F+B performance of UltraEP's plans, with the gap narrowing as workload size increases. At 1K and 4K tokens/rank, our reduction-aware planner chooses not to create replicas, whereas UltraEP creates five to eight in total. This avoids replica-weight transfers and gradient reductions, whose costs are harder to amortize at small workloads.
Finally, comparing planner execution alone, our on-device planner is $1.31$--$2.35\times$ faster than UltraEP's planner with CUDA-graph replay. Together, these results show that the observed plan quality is not specific to the MegaFlux executor, while our planner incurs lower planning overhead.


\WFclear
\subsection{End-to-End Evaluation and Discussion}

\newsavebox{\vllmTableBox}
\sbox{\vllmTableBox}{%
  \normalsize%
  \setlength{\tabcolsep}{3pt}%
  \renewcommand{\arraystretch}{0.9}%
  \begin{tabular}{@{}lcc@{}}
    \toprule
    Batch & 16K chunk & 32K chunk \\
    \midrule
    8  & $1.128\times$ & $1.133\times$ \\
    16 & $1.227\times$ & $1.238\times$ \\
    32 & $1.165\times$ & $1.260\times$ \\
    \bottomrule
  \end{tabular}%
}
\makeatletter\@afterindenttrue\makeatother
\begin{wraptable}{R}{\wd\vllmTableBox}
  \centering
  \normalsize
  \setlength{\abovecaptionskip}{0pt}
  \setlength{\belowcaptionskip}{2pt}
  \caption{\raggedright End-to-end prefill speedup over fixed placement.}
  \label{tab:eval-vllm}
  \usebox{\vllmTableBox}
\end{wraptable}

\Paragraph{End-to-end inference.}
We integrate MegaFlux into vLLM 0.27.1 for DeepSeek-V4-Pro-0813 on eight
B200 GPUs (EP8), using 160 LongBench-Pro prompts~\citep{chen2026longbench}, batches of 8/16/32, 16K/32K prefill chunk budgets per GPU, and two replica slots per rank.
We measure batch latency until every request produces one output token; MegaFlux yields $1.13$--$1.26\times$ median end-to-end speedups over fixed placement (\Tbl{\ref{tab:eval-vllm}}).

\Paragraph{Applicability and limitations.}
MegaFlux is most effective when routing creates substantial work imbalance
and sufficient expert computation exists to amortize planning and replica
operations.
Balanced workloads remain near parity, while small skewed backward workloads
can regress to $0.91\times$.
A runtime gate could therefore enable replication only when predicted
straggler-time savings exceed planning and exposed transfer/reduction costs;
we leave such gating to future work.
Our evaluation is limited to a single NVLink domain; topology-aware,
cross-node replication is also left to future work.
\FloatBarrier


\section{Related Work}
\label{sec:related-work}

\Paragraph{Megakernels and fused MoE execution.}
FLUX and COMET overlap distributed communication with computation through fine-grained fusion and scheduling~\citep{chang2024flux,comet}. FlashMoE, DeepGEMM's Mega MoE, CuTeDSL MegaMoE, UniEP, and Mixture-of-Kittens further fuse MoE communication and expert computation into persistent or megakernel execution~\citep{flashmoe,deepgemm,cutedslmegamoe,zheng2026uniep,mok}; MPK provides a more general in-kernel task runtime~\citep{mpk}. These systems optimize execution for a given expert placement. MegaFlux builds on CuTeDSL MegaMoE and couples persistent execution with runtime expert replication.

\Paragraph{Runtime expert replication and load balancing.}
FasterMoE and FlexMoE adapt expert replication or placement to workload skew~\citep{fastermoe,flexmoe}, while EPLB, LPLB, and CRAFT optimize replica placement or work assignment under resource constraints~\citep{eplb,lplb,zhao2026craft}. 
LLEP, MoonEP, and UltraEP react to post-gating loads by moving expert state and redistributing routed work~\citep{llep,moonep,ultraep}; MoonEP and UltraEP also aggregate gradients from replicas during training.
Related planners increasingly model execution cost beyond token counts: METRO accounts for activated experts, while TEMPO models weight streaming and tile-rounded computation~\citep{metro,tempo}.
Our focus is not replication alone, but integrating the resulting weight-transfer and gradient-consolidation dependencies into persistent expert computation with a token-block-granular planner.

\Paragraph{Overlapping the cost of rebalancing.}
Libra, FEPLB, and PROBE overlap balancing overhead using predictive prefetching, copy engines, or lookahead pipelines~\citep{libra,feplb,zhu2026probe}; PROBE additionally jointly plans replicas and token assignments subject to a transfer-hiding window. MegaFlux instead plans from the current routing and integrates replica operations into the persistent MoE schedule itself: weight availability gates forward progress, while backward gradient consolidation proceeds concurrently with expert computation.
\section{Conclusion}
\label{sec:conclusion}

We present MegaFlux, which resolves GPU stragglers due to work imbalance in persistent MoE execution via dynamic expert replication.
An on-device planner assigns replicas and token blocks, 
while the forward and backward megakernels pipeline replica-weight transfers and gradient reductions with expert computation.
Across diverse routing skew and workload sizes on eight NVIDIA B200 GPUs, MegaFlux
achieves geometric-mean speedups of $1.45\times$ in forward and $1.28\times$
in backward execution over fixed placement,  peaking at $2.14\times$ and $2.64\times$. 
This shows that persistent MoE execution can adapt to routing skew without being constrained by fixed expert placement.


\bibliography{iclr2027_conference}
\bibliographystyle{iclr2027_conference}
\appendix
\FloatBarrier

\section{Planner Details}
\label{app:planner}

\subsection{Greedy Placement and Refinement}
\label{app:planner-search}

\Paragraph{Candidate construction.}
Each search starts from the no-replica assignment $q^0_{er}=b_e\mathbf{1}[r=h(e)]$ and targets $\tau=\lceil\sum_e b_e/P\rceil$ blocks per GPU.
The four candidates combine donor priority (decreasing total overload or largest remaining home-expert load) with fan-out (one new copy per expert per round, or multiple copies).

A donor first uses its existing replicas, ordered by receiver capacity at visit entry, then considers its remaining expert allocations in decreasing size.
A new replica goes to the least-loaded GPU below $\tau$ that has a free replica slot and does not already hold the expert.
For either an existing or a new replica of expert $e$ on GPU $r$, we move
\begin{equation}
  v=\min\!\left(q_{e,h(e)},\max(0,\tau-\ell_r)\right)
  \label{eq:planner-transfer}
\end{equation}
blocks from the owner, updating allocations and GPU loads after each move. We do not cap $v$ by the donor's excess above $\tau$; the donor may become underloaded and receive other experts' work in later rounds. A donor visit ends when its load reaches $\tau$ or below, or its scan finishes. Each search stops after $s+2$ rounds or a round with no movement; $s+2$ is a bounded search budget, not a convergence guarantee.

\Paragraph{Refinement.}
We select the candidate with the smallest maximum GPU load, breaking ties by replica count. We then visit replicas in increasing allocation size at pass entry and remove any whose entire allocation fits on an existing copy under the selected maximum load, preferring the owner and then the least-loaded eligible copy.
Finally, a bounded pass moves single blocks from the most-loaded GPU to the least-loaded GPU when they share an expert and their loads differ by more than one. Neither refinement opens new replicas or increases the maximum compute load.

\subsection{Reduction-aware Planning}
\label{app:planner-reduction-aware}
\label{app:planner-cost}

\Paragraph{Owner-side reduction cost.}
Each remote replica produces full $dW_{1,e}$ and $dW_{2,e}$ partials, whose sizes do not shrink with its token count. Balancing computation can therefore concentrate gradient reduction at some owners.
After load-only candidate selection and refinement, we prune using
\begin{equation}
  n_r(q)=\sum_{e:h(e)=r}\sum_{u\ne r}\mathbf{1}[q_{eu}>0],
  \qquad
  C_\lambda(q)=\max_r\bigl(\ell_r+\lambda n_r(q)\bigr).
  \label{eq:planner-owner-cost}
\end{equation}
Here $n_r$ counts remote replicas of experts owned by GPU $r$, not replicas hosted there; multiple remote copies of one expert count separately. This proxy guides pruning rather than candidate construction and does not explicitly model transfer time or overlap.

\Paragraph{Reduction-aware penalty.}
We use the heuristic penalty $\lambda=12$, corresponding to 12 token blocks of augmented owner load per remote replica. We select this value and keep it fixed. The penalty is not intended as a precise conversion between reduction and compute time. A more principled, workload-adaptive cost model that derives this penalty from measured reduction and computation costs is left to future work.

\Paragraph{Pruning rule.}
At each iteration, select the GPU $d$ with the largest $\ell_d+\lambda n_d$, and among remote replicas of experts owned by $d$, select the one with the smallest positive allocation $v$. If $v<\lambda$, remove the replica and return its blocks to $d$. The owner's augmented load changes by
\begin{equation}
  \Delta(\ell_d+\lambda n_d)=v-\lambda<0,
\end{equation}
while the receiver's load decreases by $v$ and its owner-side replica count is unchanged. All other augmented loads are unchanged, so $C_\lambda$ cannot increase.
We repeat until the selected owner has no remote replica or its smallest replica allocation is at least $\lambda$. Ties are broken by GPU ID, then by expert and receiver IDs. Since each GPU hosts at most $s$ replicas, at most $Ps$ replicas can be
removed. Finally, we revert to $q^0$ unless the pruned plan strictly improves the proxy, i.e., $C_\lambda(q)<C_\lambda(q^0)$. Unlike load-only refinement, this pass may increase the maximum compute load in exchange for lower owner-side reduction cost.
Inference omits this pruning pass. For backward, the reduction-aware plan is selected before forward and reused unchanged for backward, including replica locations and routed-assignment destinations; planning is charged once, to forward.

\Paragraph{Mapping blocks to routed assignments.}
The planner determines how many token blocks of each expert execute on each copy, but not their individual token identities.
For each expert, we deterministically order routed assignments by source GPU and local (token, top-$k$ position), partition them into blocks of at most $\beta$ rows, and assign exactly $q_{er}$ blocks to each selected GPU. This preserves at most one partial block per expert. Token identities, top-$k$ positions, and routing weights are retained for output combination, while replica gradients are reduced at the expert owner.

\section{Additional Evaluation Details}
\label{app:eval-details}

\subsection{Trace-derived Workload Construction}
\label{app:eval-workloads}

\Paragraph{Source histograms.}
Forward uses Qwen3 retrieval histograms and backward uses OLMoE training
histograms from \Sec{\ref{sec:routing-skew}}, restricted to captures with
32K input tokens per rank.
Within each source, the 5--20, 40--60, 80--95, and 97--100 percentiles of
$\kappa$ define real-low, real-median, real-high, and real-extreme.
For each band, we sort event--layer histograms by $\kappa$ and select those
nearest its 25th, 50th, and 75th percentile positions, requiring distinct
source events.
Forward events are scheduler steps; backward events are
(optimizer step, microbatch, EP group) tuples.
These three selections are fixed across expert-count and token-count sweeps.
Thus, each trace-derived configuration contains three distinct routing
realizations; skew labels are relative within each source and do not imply
matched $\kappa$ across forward and backward.

\Paragraph{Resizing and route construction.}
Let $\rho_1\ge\cdots\ge\rho_{E_0}$ be a source histogram's normalized
expert counts, and let $F$ linearly interpolate its cumulative distribution,
with $F(0)=0$ and
$F(i/E_0)=\sum_{j=1}^{i}\rho_j$.
For $E$ target experts, we set
\begin{equation}
  \widetilde{\rho}_e=F(e/E)-F((e-1)/E).
  \label{eq:histogram-resize}
\end{equation}
We scale to $PMk$ assignments, enforce the top-$k$ feasibility constraint
$c_e\le PM$, and round while preserving $\sum_e c_e=PMk$.
We then permute counts over expert IDs, with $E/P$ experts owned per GPU,
and construct top-$k$ routes realizing the counts exactly.
The three realizations use fixed seeds 17, 29, and 43 for deterministic
expert permutation and route construction.
All methods use identical routes, routing weights, and initial ownership.
We compute $\kappa$ from the constructed routes.
This procedure preserves trace-derived expert-load distributions subject to
rebinning and top-$k$ feasibility, but does not replay the original
token-level expert combinations.

\Paragraph{Aggregation.}
In \Fig{\ref{fig:eval-main}}, a configuration is an $(E,M,\text{routing family})$ tuple.
Each routing family, including balanced and Zipf controls, has three seeded
realizations.
For each realization and method, we run two fresh processes, each with five
warmup and 20 timed iterations, and take the median rank-maximum latency.
Forward adds the separately measured median planner latency; backward reuses
the saved plan without another planning charge.
We define each method's configuration latency as the median over the six
realization--run measurements and compute speedup as fixed-placement latency
divided by MegaFlux latency.

\subsection{Reduction in Assignment Imbalance}
\label{app:assignment-imbalance}

Let $N_r^{(a)}$ denote the number of valid token--expert assignments
executed on GPU $r$ under variant
$a\in\{\mathrm{fixed},\mathrm{MegaFlux}\}$, including work assigned to
owned experts and replicas.
We count valid rows rather than padded blocks or routing-weighted
assignments.
Because rebalancing preserves all routed work,
$\sum_r N_r^{(a)}=PMk$, so the per-GPU mean is $Mk$.

We measure the fraction of the busiest GPU's excess assignments removed:
\begin{equation}
  \eta =
  1 -
  \frac{\max_r N_r^{(\mathrm{MegaFlux})}-Mk}
       {\max_r N_r^{(\mathrm{fixed})}-Mk}.
  \label{eq:assignment-imbalance-reduction}
\end{equation}
Cases with zero fixed-placement excess are excluded.
We compute $\eta$ for each routing realization, average within each
setting, and report the median over the 126 nonbalanced settings per
direction.
This yields 99.6\% for forward and 98.2\% for backward.
These values measure assignment redistribution, not latency reduction.
\end{document}